\documentclass[%
 reprint,
groupedaddress,
showpacs,
 floatfix,
 amsmath,amssymb,
 aps,
 pre,
longbibliography,
superscriptaddress
]{revtex4-1}
 
\usepackage{graphicx}
\usepackage{dcolumn}
\usepackage{bm}
\usepackage{subfigure}
\usepackage{color}
\usepackage{url}
\usepackage{float} 
\usepackage[symbol]{footmisc}

\begin{document}
\preprint{APS/123-QED}

\title{Detecting climate teleconnections with Granger causality}



\author{Filipi N. Silva}
\thanks{DAVO and FNS contributed equally to this work.}
\affiliation{Indiana University Network Science Institute (IUNI), Bloomington, IN, USA}
\author{Didier A. Vega-Oliveros}
\thanks{DAVO and FNS contributed equally to this work.}
\affiliation{Center for Complex Networks and Systems Research, Luddy School of Informatics, Computing and Engineering, Indiana University, Bloomington, IN, USA}
\affiliation{Institute of Computing, University of Campinas, Campinas, SP, Brazil}
\author{Xiaoran Yan}
\affiliation{Artificial Intelligence Research Institute, Zhejiang Lab, Zhejiang, China}
\author{Alessandro Flammini}
\affiliation{Center for Complex Networks and Systems Research, Luddy School of Informatics, Computing and Engineering, Indiana University, Bloomington, IN, USA}
\author{Filippo Menczer}
\affiliation{Indiana University Network Science Institute (IUNI), Bloomington, IN, USA}
\affiliation{Center for Complex Networks and Systems Research, Luddy School of Informatics, Computing and Engineering, Indiana University, Bloomington, IN, USA}
\author{Filippo Radicchi}
\affiliation{Center for Complex Networks and Systems Research, Luddy School of Informatics, Computing and Engineering, Indiana University, Bloomington, IN, USA}
\author{Ben Kravitz}
\thanks{To whom correspondence should be addressed. E-mails: \\santo.fortunato@gmail.com, bkravitz@iu.edu}
\affiliation{Department of Earth and Atmospheric Sciences, Indiana University, Bloomington, IN, USA}
\affiliation{Atmospheric Sciences and Global Change Division, Pacific Northwest National Laboratory, Richland, WA, USA}
\author{Santo Fortunato}
\thanks{To whom correspondence should be addressed. E-mails: \\santo.fortunato@gmail.com, bkravitz@iu.edu}
\affiliation{Indiana University Network Science Institute (IUNI), Bloomington, IN, USA}
\affiliation{Center for Complex Networks and Systems Research, Luddy School of Informatics, Computing and Engineering, Indiana University, Bloomington, IN, USA}
\thanks{Test}

\date{\today}

\begin{abstract}
Climate system teleconnections are crucial for improving climate predictability, but difficult to quantify. Standard approaches to identify teleconnections are often based on correlations between time series. Here we present a novel method leveraging Granger causality, which can infer/detect relationships between any two fields. We compare teleconnections identified by correlation and Granger causality at different timescales. We find that both Granger causality and correlation consistently recover known seasonal precipitation responses to the sea surface temperature pattern associated with the El Ni\~{n}o Southern Oscillation. Such findings are robust across multiple time resolutions. In addition, we identify candidates for unexplored teleconnection responses. 
\end{abstract}

\maketitle

\section*{Introduction}

Climate system teleconnections are responses in one region of the globe to perturbations far away.  The most well-known and well-studied teleconnection stems from the El Ni\~{n}o Southern Oscillation (ENSO):  a warmer or cooler anomaly condition in the tropical Pacific Ocean that results in changes in seasonal weather patterns in (for example) North America and Asia~\cite{yuan2018}.  Other well-studied teleconnections include correlations between midlatitude temperature features that are mediated by Rossby waves~\cite{hoskinskaroly} and potential links from Arctic sea ice loss to midlatitude winter storms~\cite{cohen2014}.

In all cases, there must be physical mechanisms that link the forcing in one region to a response in another.  In the case of the non-regular and complex  ENSO fluctuations, equatorially propagating Kelvin waves trigger the event, leading to numerous downstream effects that are mediated by other modes of variability, including the Pacific Meridional Mode, the Pacific Decadal Oscillation, and the Indian Ocean Dipole~\cite{yuan2018,mcphaden2015,trenberth2019,timmerman2018,chiangvimont2004,dilorenzo,stuecker2017,chang2007,stuecker,mantuahare}.  In addition, there is evidence of ENSO teleconnections via stratospheric mechanisms~\cite{domiesen}.  Uncovering these mechanisms can be a time-consuming, research-intensive process.  Methods that have been explored in previous studies include empirical orthogonal functions~\cite{liu2002}, Green's function approaches~\cite{hillming2012,harrop2018,liu2020}, the fluctuation-dissipation theorem~\cite{fuchs2015}, and system identification.  Each of these methods has advantages and disadvantages and situations in which they are most useful.

A standard technique to detect teleconnections (prior to uncovering the underlying mechanisms) relies on computing correlations between time series of climate variables in different locations~\cite{donner_wiedermann_donges_2017}. Typically the world map is turned into a latitude-longitude grid, and each grid cell is represented by time series of variables measured in that location, e.g., temperature, precipitation, or pressure. Then, correlations between time series 
in separate locations are computed~\cite{Fan2017,yamasaki08,boers19,wang13}. Most methods in the literature rely on computing linear correlation coefficients. Nonlinear scores, like mutual information~\cite{kantz04}, do not alter the picture substantially~\cite{donges09}. The ensuing relationships of strongly correlated time series can be interpreted as the links of a \textit{climate network}, with the grid cells or locations as nodes~\cite{tsonis08,yamasaki08,donges09,steinhaeuser11,Santos2020,guez12,wang13,donner_wiedermann_donges_2017,Fan2017,boers19}. Structural features of climate networks have been associated to critical climate phenomena, like ENSO~\cite{Fan2017,yamasaki08} and Rossby waves~\cite{boers19,wang13}.

Although correlation-based methods have yielded useful results, they do not necessarily provide causal relationships among phenomena.
They also suffer from problems related to  interpretability~\cite{fentonneil}; for instance, low correlation values can lead to significant $p$-values depending on the length of the time series (for example, see Appendix A). When cross-correlation is employed, a peak in a correlogram may be spurious~\cite{damos2016using} --- for example, it can be a consequence of processes with feedback mechanisms~\cite{chatfield2019analysis}, which are common in climate science. Moreover, one can observe high computed cross-correlation peaks even among any two time series due to serial autocorrelation within each time series, which is another common feature in geophysical time series~\cite{Falasca2019,mcgraw}.

Here we explore a new method of measuring the strength of teleconnections by employing Granger causality~\cite{granger69}, a concept that has gained increasing currency in climate science~\cite{triacca,mcgraw,papagiannopoulou} and differs from standard correlation in several important ways. First, Granger causality is defined explicitly as a statistical test to assess the predictive relationship of one time series with another. Thus, it has a clear interpretation that depends only on the choice of the null models. It also satisfies the two traditional assumptions for causal relationships: given two events such that $A$ causes $B$, $A$ must happen before $B$; and $A$ must provide unique information needed to predict $B$, thus implying an information flow from $A$ to $B$. Finally, Granger causality can be extended to a multivariate framework to incorporate multiple time series into the model.

In this work we employ Granger causality to uncover potential causal relationships between temperature anomalies in the tropical Pacific and precipitation anomalies in the rest of the globe (Fig.~\ref{fig:schematic}). We show that our approach is able to recover known ENSO teleconnections and to identify candidates for new ones. Our methodology is robust and represents a  powerful new tool to investigate how remote locations influence each other's climate. 

\section*{Methods}

\begin{figure*}[ht]
\centering\includegraphics[width=1.0\linewidth]{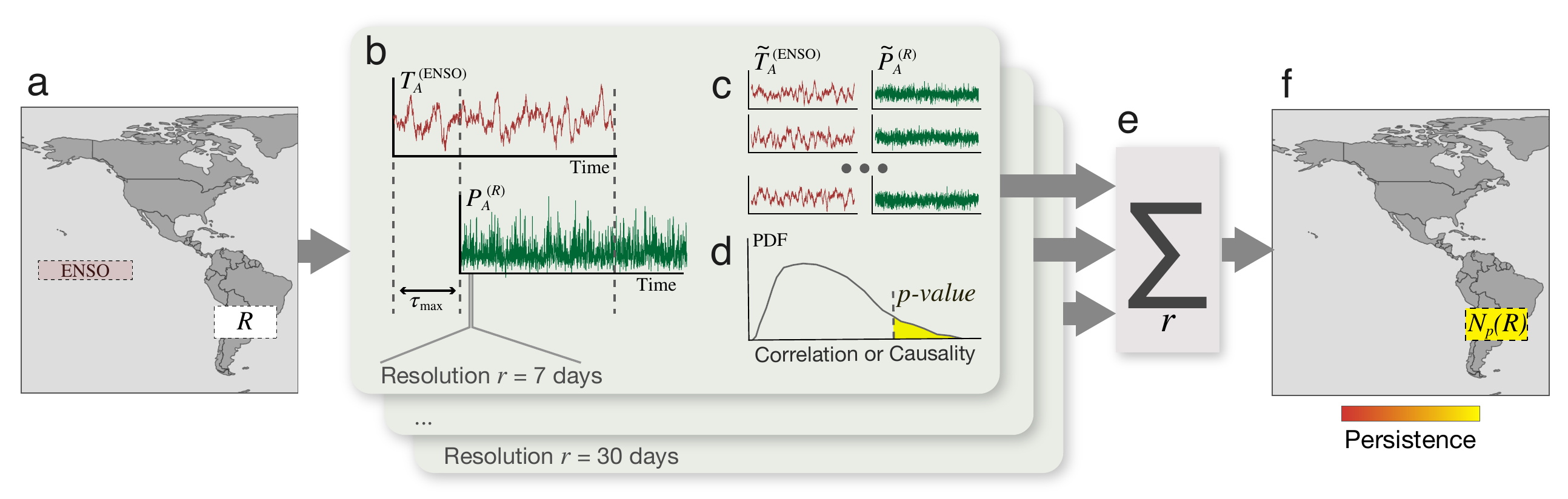}
\caption{Graphical description of the methodology used in this study (also see the Methods section). Surface air temperature data in the Ni\~{n}o~3.4 region (a) is used to perform lagged correlation or Granger causality analysis against precipitation data throughout the world, up to a maximum lag $\tau_{\mathrm{max}}$. For each considered time resolution $r$ (varying from 7 days to 30 days), we calculate the lagged correlation or Granger causality between anomalies in precipitation ($P_A^{(R)}$) in a particular region $R$ and the ENSO temperature ($T_A^{(ENSO)}$) (b). The significance of the result is found considering surrogates of both time series, i.e., $\Tilde{T}_A^{(ENSO)}$ and $\Tilde{P}_A^{(R)}$ (c), and then calculating the respective $p$-value for the $R$ region (d). For a given threshold $p_\text{thres.}$ we define that region $R$ has significant relationship with the ENSO if the corresponding $p$-value is smaller than $p_\text{thres.}$. Significant regions are aggregated and counted across the considered resolution values to measure persistence (e). Finally, colors on the maps represent regions with significant $p$-values that are persistent across $r$ (f).}
\label{fig:schematic}
\end{figure*}

The steps in the methodology, which we describe in this section, are illustrated in Fig.~\ref{fig:schematic}.

\subsection*{Dataset}

We used hourly near-surface air temperature and precipitation data from the MERRA-2 reanalysis project~\cite{Merra2}. The dataset covers the period from 1980 to 2018, with a resolution of 0.5$^{\circ}$ latitude $\times$ 0.625$^{\circ}$ longitude, for a total of 207,936 grid cells.  Of these, we only consider grid cells between $80^{\circ}$N and $80^{\circ}$S. We then reduced the spatial resolution and sample frequency by transforming the dataset to a  resolution of 1.0$^{\circ}$ $\times$ 1.25$^{\circ}$ and temporally averaging to produce  daily mean values, resulting in 365 measurements of each variable for each of the resulting 45,792 cells for each year. In the case of leap-years, we averaged the daily values of the 28th and 29th of February so all years have the same length.  Precipitation data tends to be highly skewed, with many very small values and some values orders of magnitude larger.  Therefore we re-scaled the precipitation time-series by applying a logarithmic function to all values.

We removed the seasonal cycle from the data by averaging temperature and precipitation values for each of the 365 calendar days. We considered the interval from 1 January 1980 through 28 February 2018 as the climatic period for which the long-term averages were computed.  The anomalies are then obtained by subtracting for each day the respective average temperature or precipitation from the climatic period. For example, the 1 January 1998 anomaly is computed as the value for that day minus the average of all January 1 values between 1980 and 2018.

Due to climate change over this period, the obtained anomaly time series are not sufficiently stationary in time, in that the averages in specific periods are not stable. Since our methods require approximate stationarity, we detrended the anomaly time series at each grid point by employing ordinary least squares regression of order $2$ on the time series at each point. We then subtract the fitted curve from the anomaly data. We finally smoothed the results across time by taking weekly averages of the temperature or precipitation anomalies to remove high frequency weather noise.

\subsection*{Lagged Correlation}

Cross-correlation (also sometimes called lagged correlation) methods have been commonly used in climate science to find relationships between time series~\cite{Fan2017,Falasca2019,boers19}. These techniques are based on calculating the Pearson correlation $\rho_\tau$ between two time series $x(t)$ and $y(t)$ shifted by a time lag $\tau$. One approach to find the highest correlation across all the lags is by defining an upper-bound time lag $\tau_\text{max}$ and calculating the maximum Pearson correlation across the lags as 
\begin{equation}
C_{\tau_\text{max}} = \max \{ \, | \rho_\tau \big( x(t), \, y(t+\tau) \big)  | \: : \: 0 \leq \tau \leq \tau_\text{max} \}. 
\label{eq:pearsonmax}
\end{equation}


This measure has been used as a way to construct a network of teleconnections among different regions and time series around the globe~\cite{donner_wiedermann_donges_2017}. We employ this cross-correlation as one of our methodologies to measure the relationships between temperature anomalies in the ENSO region and precipitation in each grid box.

Correlation (and cross-correlation) can be greatly influenced by the presence of outliers and by the overall distributions of the time series across the time and  frequency domains. Because of that, we employ a null-model of surrogate instances generated for each pair of the original time series 
to serve as the basis to calculate the significance of their correlations (e,.g., \cite{arizmendi}), an approach also known in statistics as bootstrapping~\cite{kantz04}. This is accomplished by calculating the $p$-value as a measure of significance among the surrogate pairs. Here, we obtained the surrogates using the refined  Amplitude Adjusted Fourier Transform (AAFT) method~\cite{theiler1992}. This technique shuffles the time series by randomizing the phases while preserving the distribution of magnitudes across the frequency spectrum. It is based on applying the Fourier transform to the data and separating the phase from the spectrum of magnitudes. Next the phase spectrum is replaced by uniformly distributed values. These surrogate data are then reassembled by means of the inverse Fourier transform. An extra iterative correction step is used to also preserve the  probability distribution of the time series, e.g., that average or extreme temperature and precipitation events are occurring with the same frequency in all realizations. Next, the $p$-value is calculated from the survival function of the surrogate distribution, more specifically, the area under the surrogate distribution curve considering all correlation values higher than $C_{\tau_\text{max}}$. 

\subsection*{Granger Causality}

The bivariate Granger causality~\cite{damos2016using,hamilton1994time} is defined as a causality test between two time series $x(t)$ and $y(t)$ according to a linear autoregressive model. If the inclusion of $y(t-\tau)$ to a linear predictive model significantly improves the prediction of $x(t)$ we say that $y$ Granger-causes (G-causes) $x$. More specifically, the test is defined in terms of the linear relationships between $x(t)$ and the lagged time series $x(t-\tau)$ and $y(t-\tau)$ for lags $\tau$ varying from $1$ to a maximum value $\tau_\text{max}$. Two linear models are taken into consideration: the \emph{complete model}, defined by
\begin{equation}
x(t) = \alpha_0 + \sum_{\tau=1}^{\tau_\text{max}} \alpha_\tau x(t-\tau) + \sum_{\tau=1}^{\tau_\text{max}} \beta_\tau y(t-\tau) + \epsilon_\text{c}(t),
\label{eq:grangerlinear}
\end{equation}
and the \emph{restricted model},
\begin{equation}
x(t) = \gamma_0 + \sum_{\tau=1}^{\tau_\text{max}} \gamma_\tau x(t-\tau) + \epsilon_\text{r}(t),
\label{eq:grangerlinearRestricted}
\end{equation}
where $\alpha_\tau$, $\beta_\tau$ and $\gamma_\tau$ are constants that can be determined from the data by using ordinary least squares, and $\epsilon_\text{c}$ and $\epsilon_\text{r}$ correspond to the residuals of the complete and restricted models respectively.

The restricted model can be regarded as a null model for the hypothesis of $x(t)$ having no dependence on $y(t-\tau)$ (i.e. $\beta_\tau = 0$ for any $\tau$) in the bivariate analysis. The performance of the two models can then be compared in terms of their residuals. Thus, $y$ G-causes $x$ ($y \rightarrow x$) when the sum of the squared residuals for the complete model $R_\text{c} = \sum_{t}[\epsilon_\text{c}(t)]^2$ is significantly smaller than those observed for the restricted model, $R_\text{r} = \sum_{t}[\epsilon_\text{r}(t)]^2$.

For the Granger causality analysis, we employed a similar pipeline to the lagged correlation analysis. However, to accelerate its computation, instead of applying the significance test based on surrogate time series, we directly performed an F-test with $\chi^2$ asymptotic approximation~\cite{hamilton1994time} based on the sum of the residuals. In this test, a $\chi^2(\tau_\text{max})$ distribution is used to approximate the null model distributions of  
\begin{equation}
F = (N - \tau_\text{max})\frac{R_\text{r} - R_\text{c}}{ R_\text{c} }  \overset{\text{null}}{\approx} \chi^2(\tau_\text{max}),
\end{equation}
where $N$ is the number of points in the time series. Finally, the $p$-values can be calculated from the survival distribution of $\chi^2(\tau_\text{max})$. As for lagged correlation, we select the more significant relationships by filtering the $p$-values in the higher percentile of the distribution. This way, we obtain  a set with the most significant locations (grid cells) impacted by ENSO.

\subsection*{Persistence}

\begin{figure*}[htb]
\centering\includegraphics[width=0.85\linewidth]{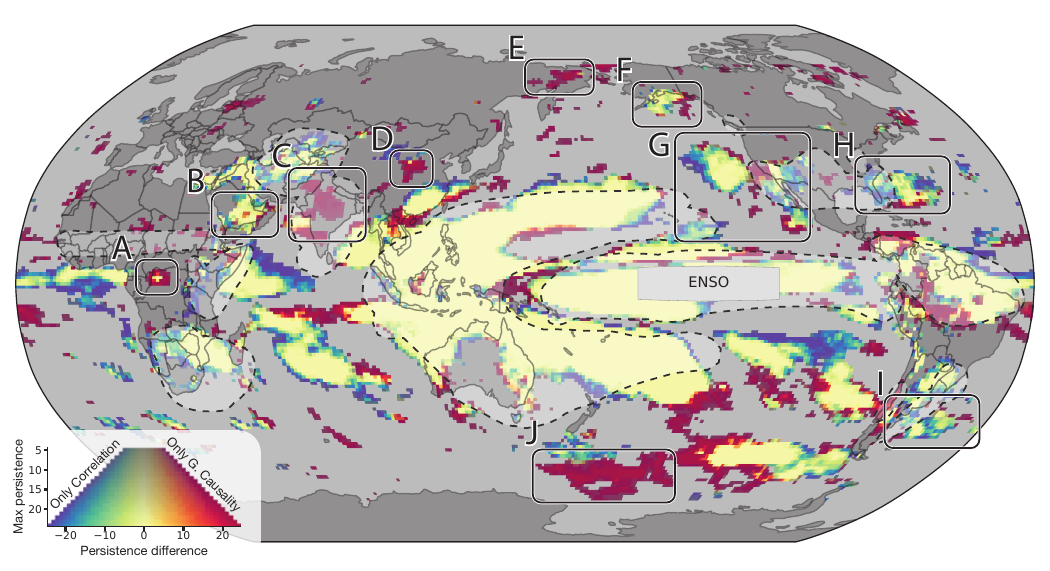}
\caption{Teleconnections between temperature in the Ni\~{n}o~3.4 region and precipitation everywhere, following the methodology described in Fig.~\ref{fig:schematic} for Granger causality and lagged correlation, each with 90-day maximum lags.  Colors indicate the persistence (see text):  differences in the number of instances (choices of time resolution between 7 and 30 days) for which there is a statistically significant result ($p\leq 0.01$) at each location, with a minimum of 5.  Yellow values indicate that lagged correlation and Granger causality are statistically significant for an equal  number of instances.  Blue values indicate that lagged correlation has more statistically significant instances than Granger causality, and vice versa for red values. The boxes inside the panels show some detected regions that are different between the methods.  Light shading indicates ground-truth regions~\cite{IRI2021a,IRI2021b}. Maps for each considered decade can be found in Appendix Fig.~A7 as well as a discussion on the major events across the decades~\cite{robock2000,slingo,medhaug} in Appendix C.}
\label{fig:maps}
\end{figure*}

It is possible to obtain spurious results that appear to be statistically significant but only for a single time resolution.  To determine the robustness of our results to changes in the resolution of the time series, we introduce a persistence metric.  For each method (lagged correlation and Granger causality), we compute at each grid point whether there is a statistically significant relationship between ENSO temperature and precipitation in that grid box, for different resolutions $r$. We considered values of $r$ ranging between $7$ and $30$ days to remove short-term weather noise while retaining a sufficient amount of data. We define persistence as the number of instances (choices of resolutions) that result in a statistically significant relationship (minimum of 5). Figure~\ref{fig:maps} compares the persistence of the two methods: yellow values indicate equal persistence for lagged correlation and Granger causality.  Blue/red values indicate higher persistence for lagged correlation/Granger causality, respectively. See Appendix A for a comparison between the Granger Causality persistence calculated using the surrogate-based null-model and the approximated distribution.

\section*{Results}

We consider as ``ground truth'' a number of regions where teleconnections have been reported in prior literature; the regions were digitized from \cite{IRI2021a,IRI2021b}, which are composites of  results beginning with \cite{Ropelewski87}.
Most of these regions (shaded areas in Fig.~\ref{fig:maps}) are well captured by both Granger causality (90-day maximum lag) and cross-correlation (also for 90-day maximum lag).  Qualitatively, the two methods perform equally well in many of the ground-truth regions, such as the Equatorial Pacific, the horseshoe-shaped pattern in the Western and Central Pacific, the Australian Coasts, the Amazon, Southern Africa, and the Horn of Africa.  Both methods tend to miss ground-truth ENSO features in the Sahel, but such relationships are variable and not robust~\cite{janicot2001}.  There are several regions (boxes in Figure~\ref{fig:maps}) where performance differs between lagged correlation and Granger causality.   (Also see Appendix A.)

The response over the Indian subcontinent (Region C) is different for the two methods, with only Granger causality recovering a statistically significant, persistent result for $p\leq 0.01$.  Indian Summer Monsoon Rainfall (ISMR) and ENSO are known to be weakly anticorrelated; the magnitude of correlation is often much less than 0.5~\cite{hrudya2020,ashok2019}.  Moreover, the correlation between ISMR and ENSO may have been weakening in recent years, possibly due to effects from increased greenhouse gas concentrations~\cite{kumar1999}, although stochastic variability could also cause an apparent weakening in correlation~\cite{yuntimmerman}.  There is little consensus about whether there is a strong relationship between ISMR and ENSO~\cite{hrudya2020}.  Nevertheless, the Granger causality method does reveal a strong relationship between Ni\~{n}o~3.4 temperature and precipitation in central India.  We hypothesize that part of the difficulty in determining whether there is a robust relationship between ISMR and ENSO is because correlation appears not to be an effective diagnostic tool in this case.

For Western North America and the nearby Eastern Pacific (Region G in Figure~\ref{fig:maps}), the picture is more mixed, with some regions showing a stronger response under lagged correlation and some under Granger causality.  While there is a long history relating El Ni\~{n}o events and precipitation in western North America, especially California~\cite{schonher,andrews2004}, as well as changes in California precipitation due to changing ENSO characteristics in response to climate change~\cite{wang2004,yoon2015}, recent studies have cast doubt on the robustness of that relationship~\cite{obrien2019}. In  \cite{kumarchen2020} the authors argue that predictability of the ENSO response in California using common methods is inhibited by low signal-to-noise ratios.  Nevertheless, Region G does show a region in southern California that is almost exclusively identified by Granger causality, potentially indicating an advantage for Granger causality in low signal-to-noise ratio regimes.  Further study is warranted.

In \cite{kumarchen2020} the authors argue that Eastern North America and the nearby Western Atlantic (Region H) are known to be influenced by ENSO in boreal winter (DJF) \cite{Ropelewski87}.  Both methods show some statistically significant teleconnections in this region, and cross-correlation has a robust response in more grid boxes.  Nevertheless, neither method recovers the robust DJF precipitation response described in \cite{kumarchen2020}.  It is difficult to pinpoint a reason behind this discrepancy.  It could be due to different performance metrics  (in \cite{kumarchen2020}, the authors used the anomaly correlation coefficient, which would not make sense to compute in our current study), as well as a different observational set; MERRA-2 is known to have substantial dry biases in this region, which could obscure teleconnections \cite{reichle2017}.  Further study is warranted before we make firm conclusions about this region.

\begin{figure*}[ht]
\centering\includegraphics[width=0.95\linewidth]{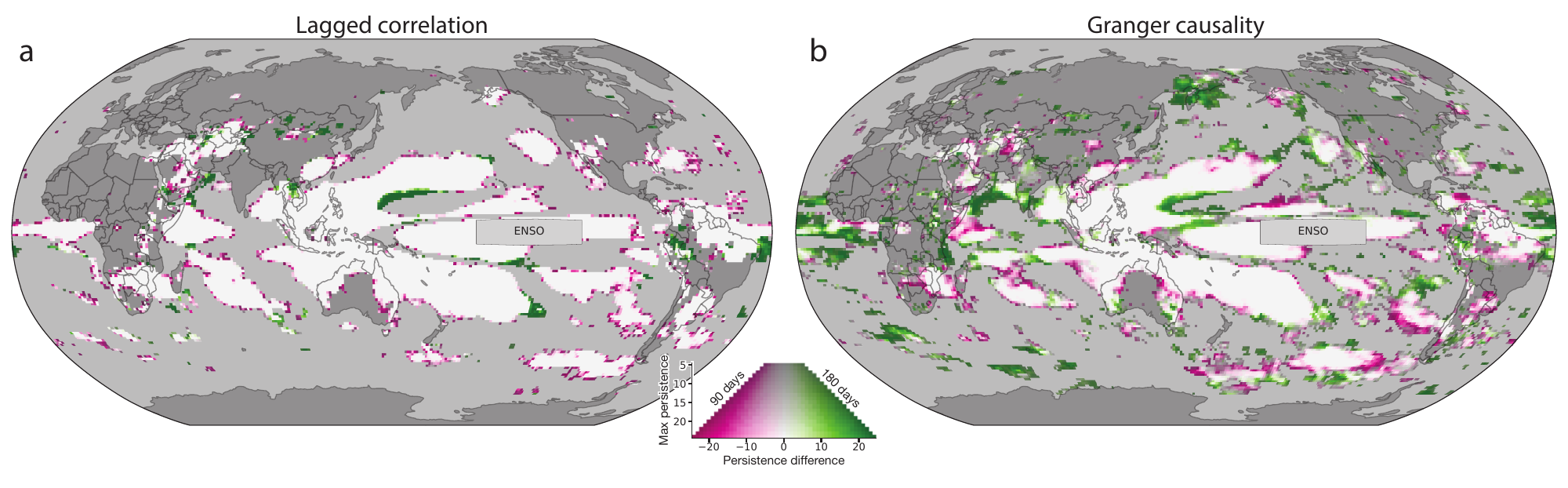}
\caption{Maps of persistence differences between maximum lags of 90 days and 180 days for lagged correlation (a) and Granger causality (b).  These maps were obtained for regions with $p < 0.01$.}
\label{fig:alllags}
\end{figure*}

It is perhaps not surprising that many regions identified by cross correlation fall within the ground truth, as many past analyses of teleconnections used correlation-based methods~\cite{masongoddard2001}.  In addition, both Granger causality and cross-correlation have identified several regions that are predominantly picked up by only one method.  Regions B and I in Fig.~\ref{fig:maps} are more persistent with cross-correlation but do not substantially overlap with the ground truth.  For these regions, Granger causality does detect at least some portion of the areas that the correlation-based method does, but for different maximum lag windows (Fig.~\ref{fig:alllags}b), indicating that the maximum lag window is an important parameter in this methodology. Conversely, the maximum lag window does not appear to be important for cross-correlation (Fig.~\ref{fig:alllags}a), which is counter-intuitive: there should be evolving seasonal dynamics of precipitation. This suggests a potential shortcoming of the cross-correlation method; further discussion of this issue can be found below.

Several regions appear to be persistent only for Granger causality (Fig.~\ref{fig:maps}), and with few exceptions, do not appear in the ground truth.  We are reluctant to make any conclusions about whether the regions indicate real features of ENSO teleconnections.  We highlight this as an area of future work, not only for understanding teleconnections, but also as further validation of our methodology.

It may appear strange that there are regions that satisfy Granger causality but do not appear to have a significant lagged correlation with El Ni\~{n}o~3.4 temperature (e.g., Regions C or J in Figure~\ref{fig:maps}) according to our methods.  Despite the fact that this can occur in real-world time series~\cite{altman2015}, we attribute these phenomena to specific shortcomings of lagged correlation that are overcome by Granger causality  (see Appendix A and B).  Thresholds of statistical significance for cross-correlation are sensitive to the choice of the null model, whereas that is not the case for Granger causality.  This effect can be easily demonstrated for a simple synthetic autoregressive time series (see Appendix B) that shares many characteristics with large-scale geophysical data (e,.g., \cite{macmynowski2011}): cross-correlation only chooses a single lag of maximum correlation, whereas Granger causality incorporates information across all lags (up to $\tau_{\mathrm{max}}$), resulting in a more robust metric.  Also, while the results of the two methods are often consistent, in some cases the cross-correlation method may select a value with a slightly higher correlation with a lag that is temporally inconsistent with neighboring grid points, which is a physically spurious result. 

\begin{figure}[ht]
\centering\includegraphics[width=0.95\linewidth]{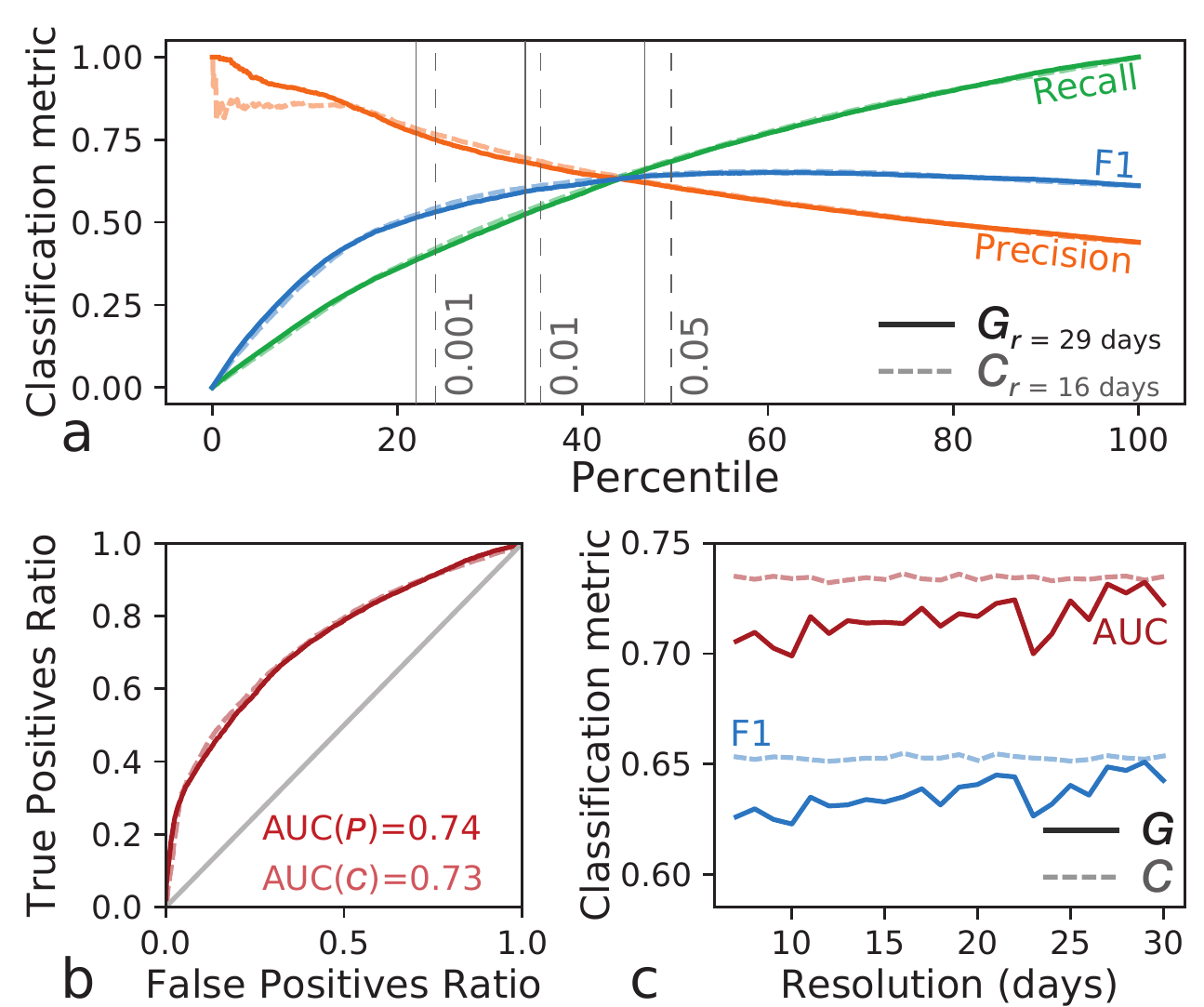}
\caption{Classification metrics for lagged correlation (solid lines) and Granger causality (dashed lines). (a)~Precision, recall and F1 score vs. percentile of the top-most significant relationships, with vertical lines indicating the percentile positions corresponding to $p$-values 0.05, 0.01, and 0.001. (b)~Receiver operating characteristic (ROC) curves. The value of the area under the curve (AUC) is shown for both methodologies.  (c)-The AUC and maximum F1 scores for Granger causality (solid) and lagged correlation (dashed) as a function of the time resolution in days.  See Appendix D for further description of these metrics.}
\label{fig:fmeasuresPG}
\end{figure}

Figure~\ref{fig:fmeasuresPG} shows globally aggregated metrics gauging how well cross-correlation ($C_\tau$) and Granger causality ($G_\tau$) perform against the ground truth for a given maximum lag $\tau$ (in days). For this comparison we adopted only the best results of each model, corresponding to $C_{16}$ and $G_{29}$ (a more complete set of plots is available in Appendix, Fig.~A7). While the curves are nearly identical, Granger causality has slightly worse quantitative performance than cross-correlation as measured by the AUC score. This is likely because most of the errors in cross-correlation result from false negatives (misses), whereas in addition to false negatives, Granger causality also identifies several  features (e.g., regions D, E, and J in Fig.~\ref{fig:maps}) that do not appear in the ground truth, which register as false positives. This effect is more clear by looking at precision and recall metrics (Appendix D). Granger causality displays higher precision and recall than cross-correlation if we keep only the areas of lower $p$-values. This indicates that the most significant Granger causal relationships occur inside the ground-truth regions, lending confidence that the lower AUC scores for Granger causality are indeed primarily because that method identifies new features that are not in the ground truth.

To assess the similarity of the maps obtained by employing the two methodologies, we computed the Spearman rank correlation between $C_{\tau_\text{max}}$ (cross-correlation) and $G_{\tau_\text{max}}$ (Granger causality) for different values of the maximum time lag $\tau_\text{max}$ (see Table~A1). There is not much difference in the ranks for the cross-correlation method across the considered values of $\tau_\text{max}$ (rank correlation values range between 0.77 and 0.94). On the other hand, Granger causality seems to be more sensitive to the maximum time lag because the autoregressive model encompasses all the information of previous lags (up to $\tau_\text{max}$) to predict the time series.

\section*{Discussion and Conclusions}

We have developed a new method of quantifying climate system teleconnections using Granger causality that complements existing methods like cross-correlation.  In addition to accurately reproducing the results found in the ground truth (as cross-correlation also does), this new method has several advantages:

\begin{itemize}
    \item Granger causality identifies statistically causal relationships with high precision and recall when focusing on high-confidence regions (low $p$; Fig.~\ref{fig:fmeasuresPG}).
    
    \item Cross-correlation appears to not vary substantially with the maximum lag window.  Based on the information in Fig.~\ref{fig:alllags}, we cannot conclude whether cross-correlation is indeed picking up longer-term modes of variability, as we would want for evaluating ENSO teleconnections.  Conversely, relationships found by Granger causality differ with lag (Fig.~\ref{fig:alllags} and Appendix B), which not only matches our physical intuition that relationships evolve over time, but also allows us to determine the timescale of different precipitation responses to Ni\~{n}o 3.4 temperature and, in principle, the spatiotemporal evolution of teleconnections.
    
    \item Cross-correlation methods can cause spurious results (e,.g., \cite{mcgraw}), appearing as false positives due to serial autocorrelation among the time series. As was seen for the Indian Summer Monsoon, there is also potential for false negatives due to a relative insensitivity. We also encountered a U-shaped distortion for the distribution of maximum lags~\cite{yule_why_1926,dean_dangers_2016}, which depends on the window size (see Fig.~A5 in the Appendix, as well as an extended discussion in Appendix A). Thus the selected best lags for cross-correlation, in some cases, are artifacts of the method.
\end{itemize}

The potentially novel results highlighted by the regions in Fig.~\ref{fig:maps} need further investigation.  Our method is useful for discovering statistically robust relationships, but these explanations are incomplete without connecting physical mechanisms. To aid in this process, one promising attribute of Granger causality is the ability to construct long causal chains, that is, the discovered relationships indicated in Fig.~\ref{fig:maps} may be mediated by several intermediate steps.  Other proposed techniques can detect indirect relationships or retrieve the causal graph (e,.g., \cite{zhou2015teleconnection,runge2019detecting}). Future work could explore indirect associations through a multivariate Granger causality framework, which is more suited to be compared to the mentioned techniques used to retrieve causal graphs.  This could be useful for uncovering physical mechanisms and signal propagation through the Earth system.

Several aspects of our method require further exploration.  An example is the maximum lag --- for Granger causality, this appears to be an important parameter.  We have not yet explored the ability of this parameter to reveal useful properties of relationships, such as signal propagation.  Moreover, the lags we chose had an upper limit of 180 days, which is a reasonable limit for many of the effects of ENSO. However, that upper limit may need to be altered for other applications, which could  introduce difficulties in obtaining clear signals, such as if the annual cycle interferes with the calculations or if other modes of variability begin to play important roles; these potential effects have not been explored.  Other aspects of our method that need to be investigated include time averaging or filtering the precipitation results by percentile.  These potential effects may impact the robustness of our method, but we have not explored them in detail. 

Perhaps one of the most important aspects of our method is that it provides additional testable predictions about thus far unverified ENSO teleconnection regions.  The ground truth we used for ENSO teleconnections (shaded regions in Fig.~\ref{fig:maps}) is a composite of results obtained over years of research.  Our methodology has identified new regions, which need to be reconciled with that composite, resulting in rejection of our findings or an update to the suite of known ENSO teleconnection responses.  This process could apply to teleconnection studies more generally, providing a systematic way of identifying teleconnections between any two climate fields; this could be especially useful for features in which the response is difficult to understand or simply unknown, for example, for attribution of changes in extreme weather event frequency.

\begin{acknowledgments}
This research was developed using the computational resources from the Center for Mathematical Sciences Applied to Industry (CeMEAI) funded by the Fundação de Amparo à Pesquisa do Estado de São Paulo (FAPESP) under Grant No. 2013/07375-0. D.A.V.O acknowledges FAPESP Grants 2016/23698-1, 2018/24260-5, and 2019/26283-5. This material is based upon work supported by the National Science Foundation under Grant No. CNS-0521433.  Support for B.K. was provided in part by the National Science Foundation through agreement CBET-1931641, the Indiana University Environmental Resilience Institute, and the {\it Prepared for Environmental Change} Grand Challenge initiative.  This work was based on research supported by the U.S. Department of Energy (DOE), Office of Science, Biological and Environmental Research, as part of the Regional and Global Model Analysis program.  The Pacific Northwest National Laboratory is operated for the US Department of Energy by Battelle Memorial Institute under contract DE-AC05-76RL01830. \\
\end{acknowledgments}

Code to reproduce the analysis presented in this manuscript is available at \\ \url{https://doi.org/10.5281/zenodo.5266377}

\bibliography{climate}

\clearpage

\appendix

\renewcommand{\thefigure}{A\arabic{figure}}
\renewcommand{\thetable}{A\arabic{table}}
\setcounter{figure}{0}
\setcounter{table}{0}

\section{Statistical significance for Granger causality and lagged correlation}

In the main text, we referred to numerous differences between Granger causality and cross-correlation, including how results can demonstrate causality but not correlation.  We expand upon those reasons here.

\begin{figure}[htb]
\centering\includegraphics[width=0.95\linewidth]{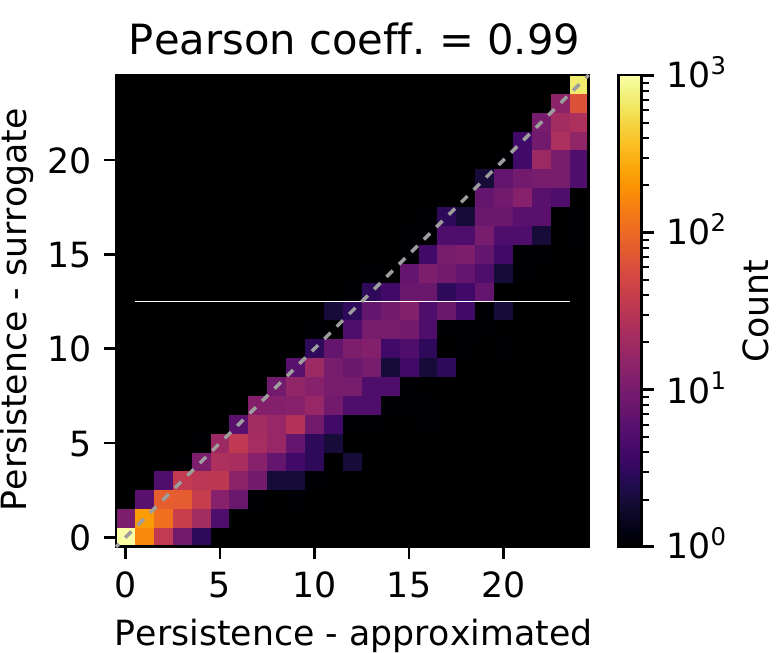}
\caption{Relationship between the approximated and surrogate-based null models to calculate the Granger Causality $p$-values. The x-axis corresponds to the persistence obtained from the approximated null model of causality defined in terms of $\chi^2$, and the y-axis corresponds to the persistence obtained from p-values calculated from $10000$ surrogate null-model realizations.}
\label{fig:persistenceCausalitySurrogateApproximation}
\end{figure}

We first compare the persistence values obtained for Granger Causality from surrogate time-series null model against those from the null-model approximation described in the paper (Figure~\ref{fig:persistenceCausalitySurrogateApproximation}). This shows that there is a strong correspondence (in terms of Pearson correlation) between the two null models. Also, regions of extreme persistence values (0 or 24) matched for both models.

\begin{figure}[htb]
\centering\includegraphics[width=0.75\linewidth]{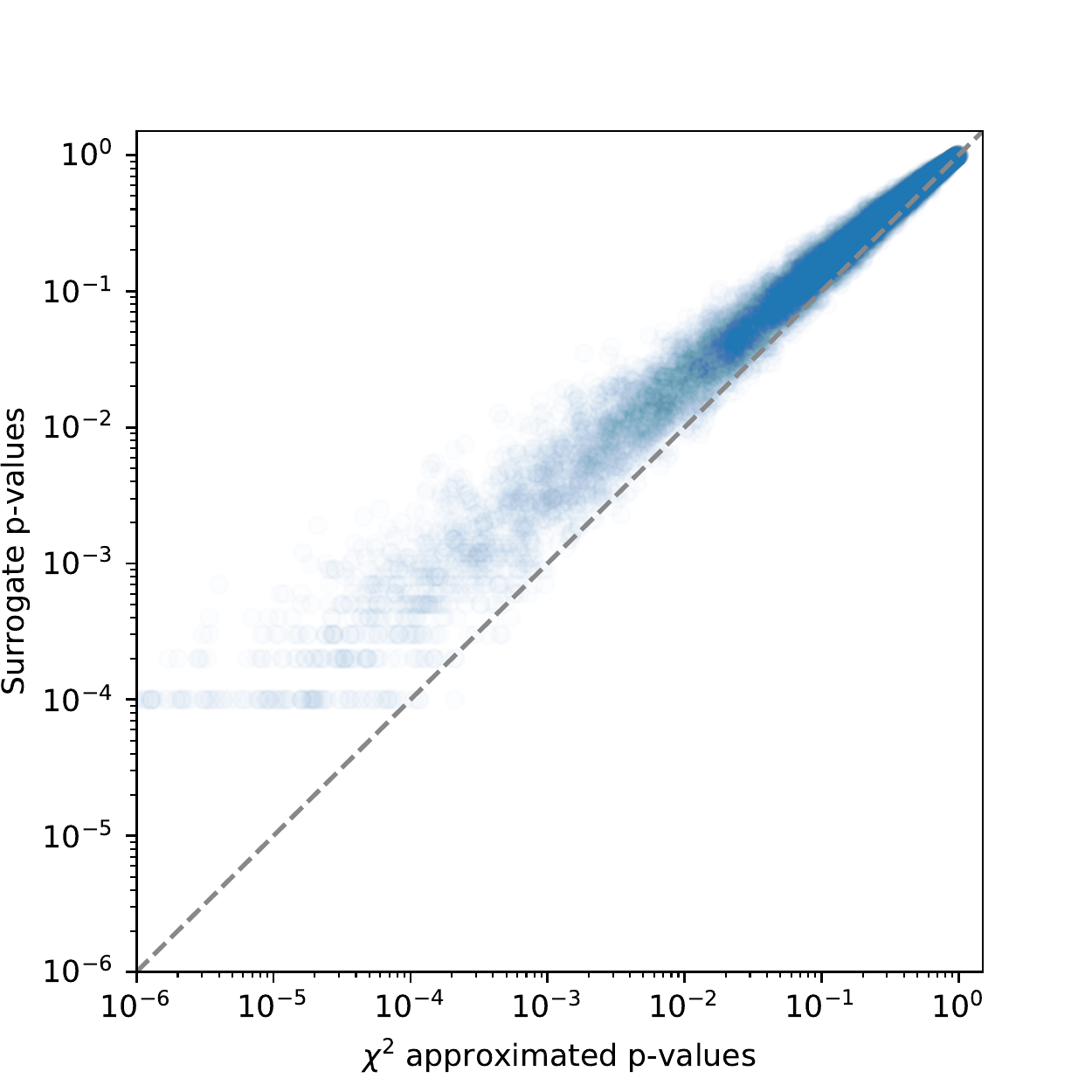}
\caption{Direct relationship between the $p$-values of the pairwise and approximated null models. The x-axis corresponds to the approximated causality defined in terms of $\chi^2$, and the y-axis corresponds to the p-value calculated from $10000$ surrogate null-model realizations.}
\label{fig:causalitySurrogateApproximation}
\end{figure}

\begin{figure}[htb]
\centering\includegraphics[width=0.85\linewidth]{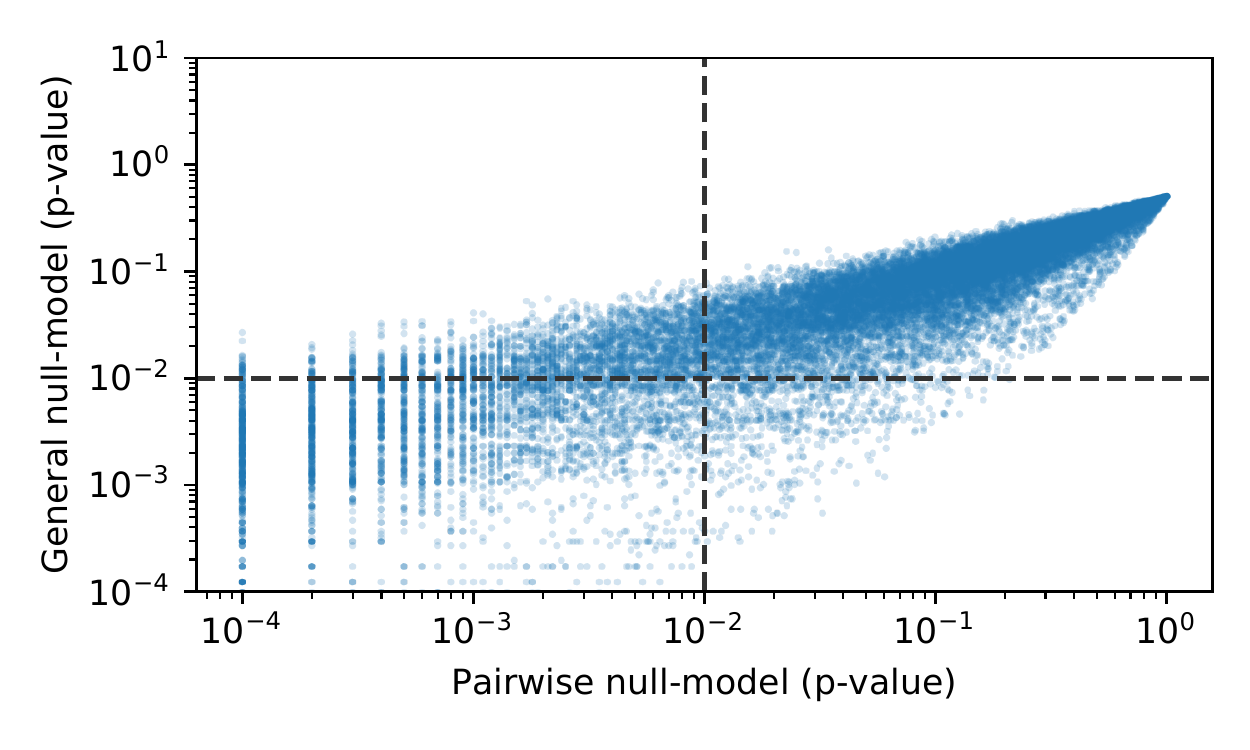}
\caption{Relationship between the general and pairwise null models for calculating lagged correlation $p$-values.  $p$-values of 0.01 are indicated by dashed lines.}
\label{fig:correlationGeneral}
\end{figure}

We also checked the correspondence between the $p$-values obtained a $7$ day window obtained for the two considered null models (pairwise surrogate and approximated), as shown in~\ref{fig:causalitySurrogateApproximation}. We use the same surrogate approach as applied to calculate lagged correlation. Because of the high computational cost, we uniformly sampled about $1/4$ of the time series for this analysis.  We found that the surrogate-based $p$-values are slightly greater than the approximated $p$-values, but the two have a strong relationship leading to a Spearman rank correlation of $0.98$.  This indicates that statistical significance using Granger causality is not sensitive to the choice of how $p$-values are calculated. Conversely, Figure~\ref{fig:correlationGeneral} shows that this is not the case for lagged correlation and that for a $p$-value threshold of 0.01 (dashed lines).  For lagged correlation, it is entirely feasible to get results that either are or are not statistically significant, depending upon the null model used. For this comparison, we employed a general null that differs from the pairwise version used in the main text. Instead of using a null model distribution for each pair of time series, the general null model distribution is drawn a single time from a set of surrogate time series obtained from the real data and used to calculate all $p$-values. 

\begin{figure}[htb]
\centering\includegraphics[width=0.85\linewidth]{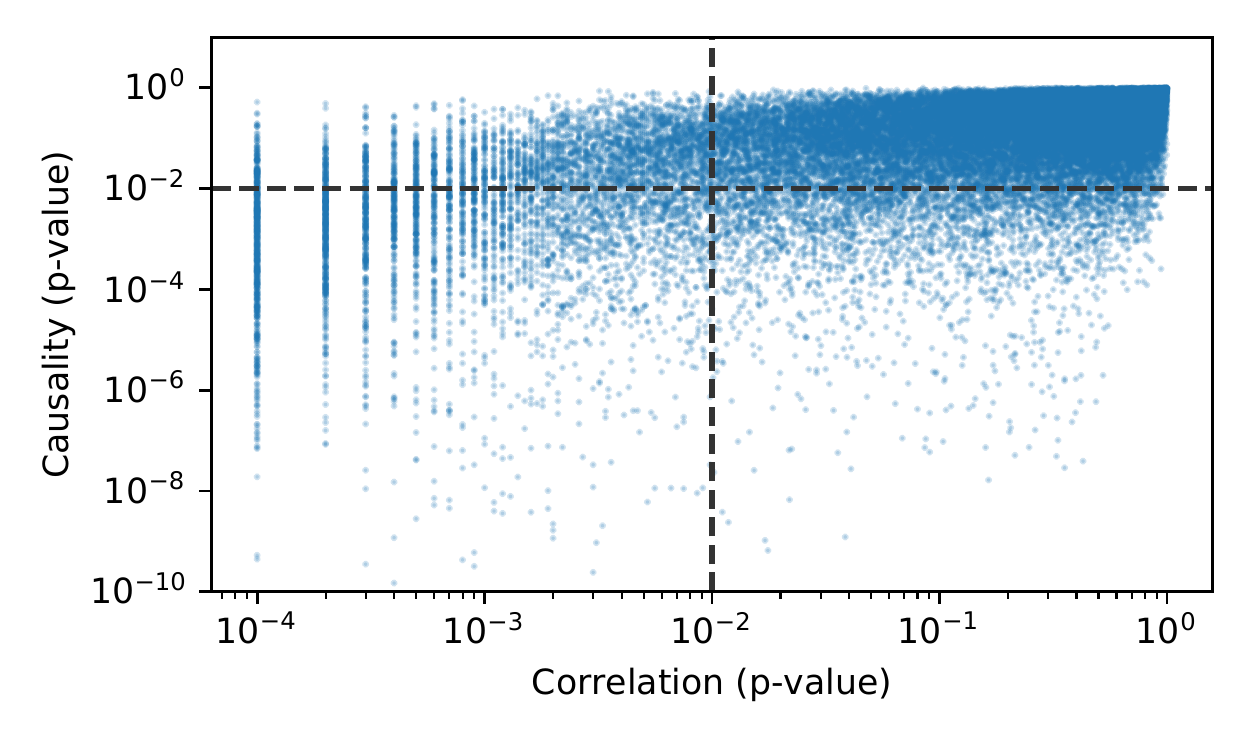}
\caption{Causality vs lagged correlation p-values.  $p$-values of 0.01 are indicated by dashed lines.}
\label{fig:correlationVsCausality}
\end{figure}

Moreover, the $p$-values obtained for the Granger causality method and the lagged correlation method are not strongly correlated (Figure~\ref{fig:correlationVsCausality}), providing further examples of how features can be significant under Granger causality but not significant under cross-correlation.

\begin{figure}[htb]
\centering\includegraphics[width=0.85\linewidth]{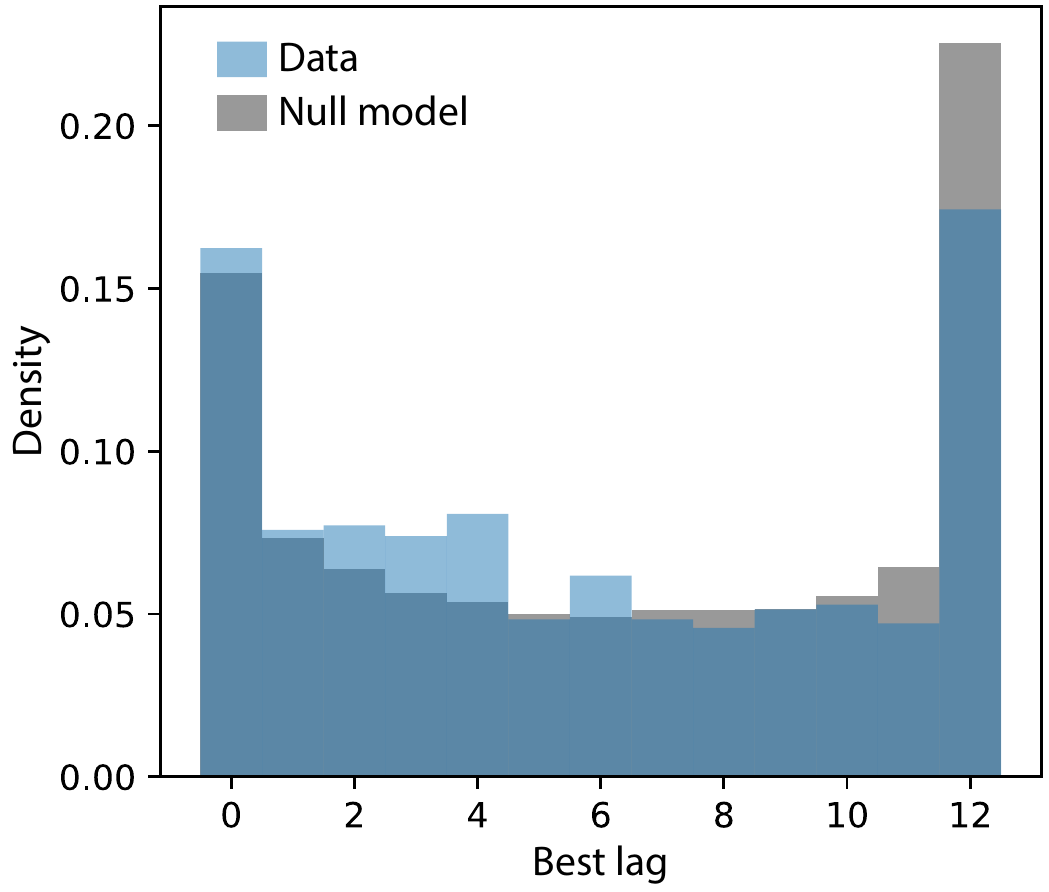}
\caption{Distribution of the best lags obtained for lagged correlation. In blue, the distribution for the data and in gray, the distribution for the null-model.}
\label{fig:bestlags}
\end{figure}

In Figure~\ref{fig:bestlags}, we show how the lags that yield the best results for cross-correlation are distributed across all the considered time series. We found U-shaped curves for both the null-model and for the data. According to the literature this is an artifact of the windowed cross-correlation method~\cite{yule_why_1926,dean_dangers_2016}. The U-shaped distortion grows stronger with more long timescale (relative to the lag window size) signals, which is widely observed in climate data.

In general, the results for Granger causality depend upon the maximum lag window, whereas the results for lagged correlation do not.  This is in large part because the maximum lagged correlation often sets up for early lags (Figures 2 and 4 in the main text, as well as Appendix Figure A6), whereas Granger causality finds the best relationship across all lags within the window.  As real signals (with associated timescales) propagate throughout the Earth system, we would expect to see variations in signal strength with lag.

Nevertheless, Figure 3 in the main text displays some curious behavior, in that there are pink regions for Granger causality, indicating that shorter maximum lag windows result in better prediction.  Intuitively, adding longer windows should always improve the predictive power of Granger causality. These results can be explained via the definition of Granger causality and the way it is computed.  For a given time series $X$ and $Y$, Granger causality $X\to Y$ measures how significant the information in X is for predicting Y {\it in comparison to predicting it only using previous values from itself}.  When we add more information, we are adding new points for both $X$ and $Y$, so if $Y$ starts to make significantly better predictions than $X$, we may see pink regions in the map in Figure 3.  This spurious behavior only happens along the margins of the shaded regions in Figure 3, which are also regions of lower statistical significance.


\section{Relationship between lagged correlation and Granger causality}
We found that Granger causality and lagged correlation are not strongly correlated. This is shown in Figure~\ref{fig:correlationVsCausality}. A table aggregating the Spearman rank correlations between the two techniques for different time lags is shown in Table~\ref{tab:spearman}

\begin{table}
\centering
\caption{Spearman rank correlations among the $p$-values of the Pearson correlation ($P_{0} = \rho$), lagged correlation ($C_{\tau_\mathrm{max}}$) and Granger causality ($G_{\tau_\mathrm{max}}$), considering ${\tau_\mathrm{max}}$ maximum lag windows of 30, 90, and 180 days.\label{tab:spearman}}
\begin{tabular}{l|ccccccc}
  & $C_{0}$ & $C_{30}$ & $C_{90}$ & $C_{180}$ & $G_{30}$ & $G_{90}$ & $G_{180}$ \\ \hline
$P_{0}$                 & 1.00  & 0.94   & 0.86   & 0.77    & 0.69   & 0.55   & 0.48    \\
$P_{30}$                &   & 1.00   & 0.94   & 0.83    & 0.67   & 0.53   & 0.46    \\
$P_{90}$                &   &    & 1.00   & 0.91    & 0.64   & 0.52   & 0.46    \\
$P_{180}$               &   &    &    & 1.00    & 0.62   & 0.52   & 0.46    \\
$G_{30}$                &   &    &    &     & 1.00   & 0.78   & 0.66    \\
$G_{90}$                &   &    &    &     &    & 1.00   & 0.81    \\
$G_{180}$               &   &    &    &     &    &    & 1.00   
\end{tabular}
\end{table}

As an additional illustration of potential reasons that can explain these figures, we constructed a synthetic autoregressive time series, described by
\begin{equation}
    Y'(t) = Y(t) + \sum\limits_{l=1}^{L_{\mathrm{max}}} \beta \frac{X(t-l)}{L_{\mathrm{max}}},
\end{equation}
$X$ and $Y$ are two time series derived from the real distributions of Ni\~{n}o~3.4 temperature and a location's precipitation, with their values shuffled.  We then added lagged relationships to $Y$ as above, resulting in $Y'$, where $L_{\mathrm{max}}$ is the maximum lag (the order of the autoregressive process), and $\beta$ is a weight parameter.  While simple, autoregressive processes are reasonable simplifications for a variety of geophysical time series, in particular that many geophysical time series have serial autocorrelation (MacMynowski et al., 2011).

\begin{figure}[htb]
\centering\includegraphics[width=0.95\linewidth]{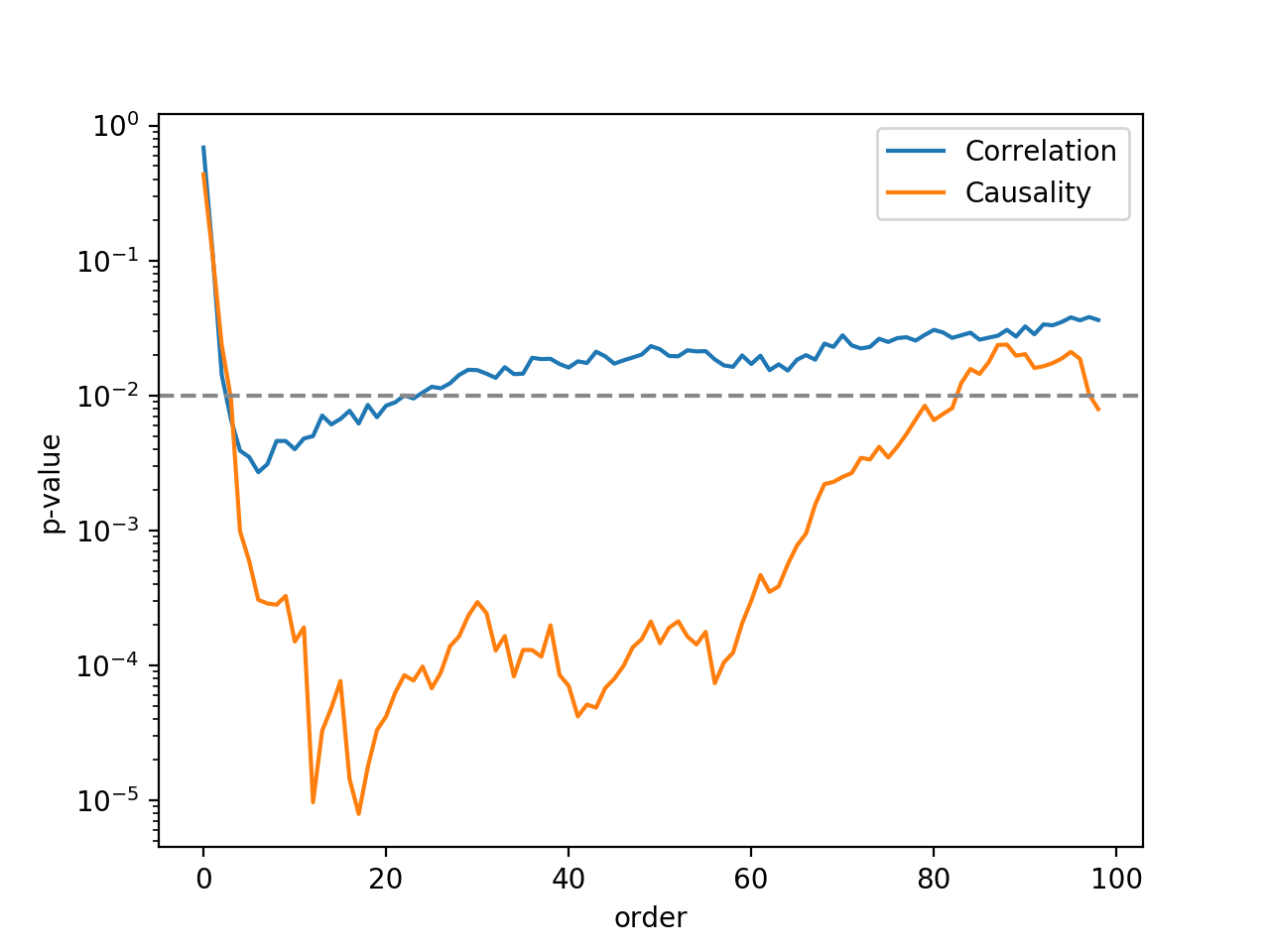}
\caption{$p$-value as a function of the order of the autoregressive model for lagged correlation (blue) and Granger causality (orange).  A p-value of 0.01 is indicated by the dashed line.}
\label{fig:toyModel}
\end{figure}

Figure~\ref{fig:toyModel} shows the results of applying both the Granger causality method and the lagged correlation method to these time series, varying the maximum lag (order).  From this figure, except for very short lags, cross-correlation is relatively insensitive to the maximum lag, and the relationship between $X$ and $Y'$ is not statistically significant for many lags (p=0.01).  Conversely, because Granger causality is able to take all lags in the window $[0,L_{\mathrm{max}}]$ into account when computing the relationships between $X$ and $Y'$, a statistically significant relationship is often found.



\section{Results by decade}
\begin{figure*}[htbp]
\centering\includegraphics[width=0.95\linewidth]{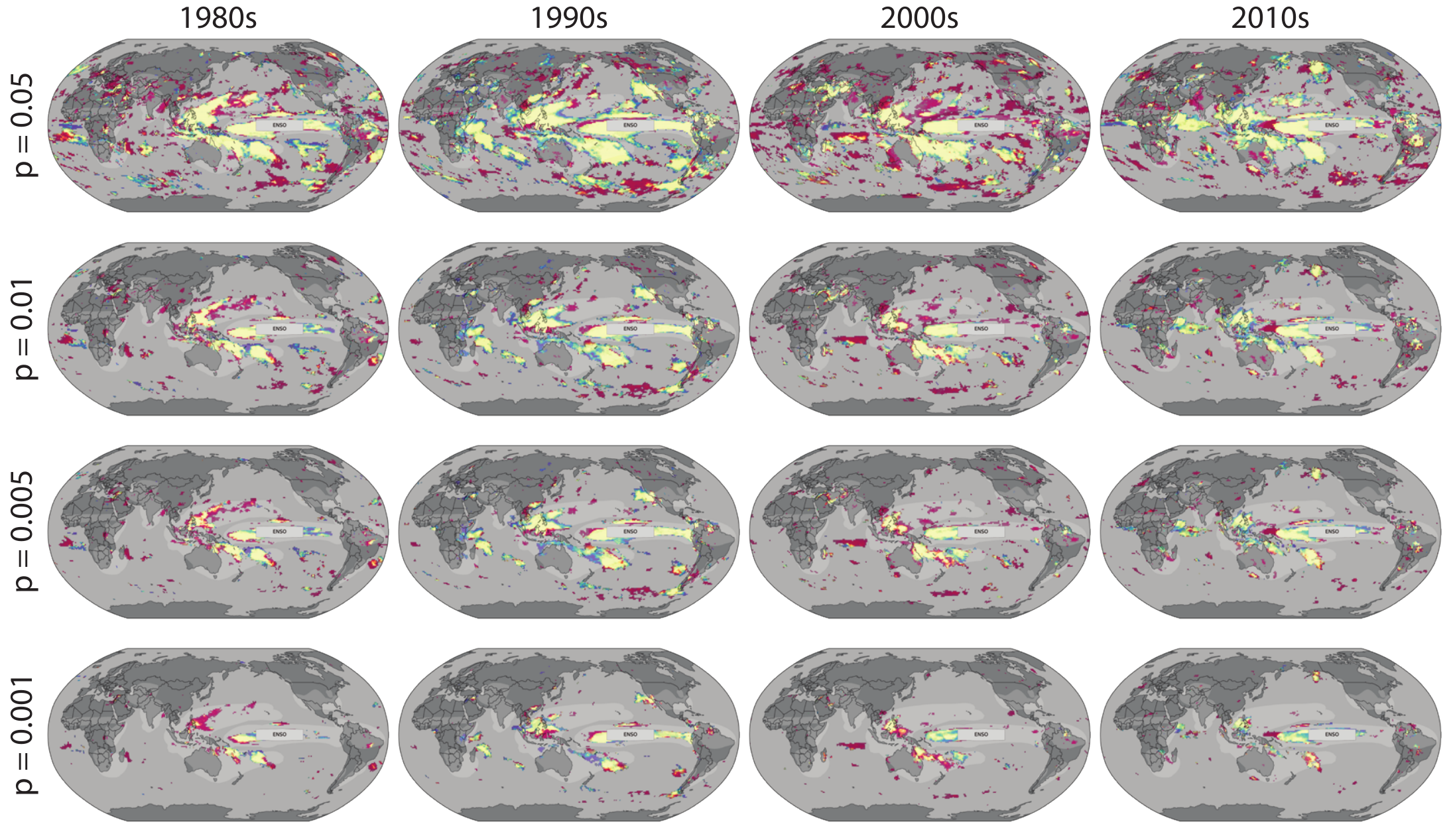}
\caption{Analogs to Figure 2 in the main text, but computed for different decades and at different $p$ values, to understand the sensitivity of the results to the amount of data used in calculations.}
\label{fig:decade}
\end{figure*}

Figure~\ref{fig:decade} shows analogs to Figure 2 in the main text, but computed for different decades and at different $p$ values, to understand the sensitivity of the results to the amount of data used in calculations.  The results shown in Figure~\ref{fig:decade} are quite noisy, and determining robust features is difficult.  This is perhaps unsurprising.  The 1980s experienced the large volcanic eruption of El Chich\'{o}n \cite{robock2000} concurrent with a large El Ni\~{n}o \cite{slingo}, as well as the least amount of global warming as compared to the other decades.  The 1990s experienced the eruption of Mt. Pinatubo in 1991, which was the largest eruption of the late 20th century \cite{robock2000}, followed by one of the largest El Ni\~{n}o events on record in 1997-1998 \cite{slingo}.  The 2000s and 2010s then experienced various levels of global warming, including phases of the ``warming hiatus'' that were likely due to modes of internal climate variability \cite{medhaug}.  All of these different compounding factors would make it difficult for robust ENSO signals to rise to the point of detection, particularly for low $p$ values.

\section{Performance Metrics}
In the main text, we refer to various metrics (Figure 4) like precision, recall, F1-score, receiver operating characteristic (ROC), and area under the curve (AUC).  It may be useful to readers for us to define them here.  For all of these metrics, TP = true positives, FP = false positives, TN = true negatives, and FN = false negatives.

\begin{equation}
    \mathrm{Precision} = \frac{\mathrm{TP}}{\mathrm{TP}+\mathrm{FP}}
\end{equation}
\begin{equation}
      \mathrm{Recall} = \frac{\mathrm{TP}}{\mathrm{TP}+\mathrm{FN}}
\end{equation}
\begin{equation}
    F1\mbox{ }\mathrm{score} = 2\cdot \frac{\mathrm{precision}\cdot\mathrm{recall}}{\mathrm{precision}+\mathrm{recall}}
\end{equation}
The ROC curve plots the TP rate against the FP rate for various thresholds, indicating the probability of detection against the probability of false alarm.  The dashed 1-1 line in Figure 4 is a line of equal probability; any curve above this line indicates a model that has predictive skill that is better than chance.  As such, the area under the ROC curve, which is abbreviated AUC, is an effective way of reducing the dimension of the ROC plot to compare different models or methodologies.

\end{document}